\documentstyle[preprint,aps]{revtex}

\begin{document}
\title{NONEXTENSIVE THERMOSTATISTICAL APPROACH TO THE THERMOLUMINESCENCE DECAY}
\author{O. Kayacan$^{a,}\thanks{%
E-mail: ozhan.kayacan@bayar.edu.tr}$ and N. Can$^{a}$}
\address{$^{a}$ Department of Physics, Faculty of Art and Science, Celal Bayar\\
University, 45030\\
Muradiye, Manisa-TURKEY}
\maketitle

\begin{abstract}
In this study, thermoluminescence decay is investigated within Tsallis
thermostatistics(TT). We belive that this is the first attempt to handle
themoluminescence decay process within TT. \newline
\newline
\noindent

\end{abstract}

\newpage

\section{Introduction}

\noindent The luminescence intensity of a pre-irradiated phosphor is
recorded as a function of the rising temperature. The resulting curve
between the luminescence intensity and temperature, known as the glow curve,
first increases and then, after reaching a maximum, decreases. If n
represents the ocuupied trap density at time t, then the following equation
gives the rate of thermal release of electron at temperature T:

\begin{equation}
\frac{dn}{dt}=-ns\exp (-E/k_{B}T).
\end{equation}

If the phosphor is heated at a constant rate of increase of temperature,
i.e. $dT/dt=cons\tan t=r,$ the above equation can be rewritten as 
\begin{equation}
\frac{dn}{n}=-\frac{s}{r}\exp (-E/k_{B}T).
\end{equation}

If assuming initially at the start of heating, $T=T_{i}$, $n=n_{0}$, then
the integration of Eq.(2) results in 
\begin{equation}
n=n_{0}\exp \left( -\frac{s}{r}\int_{T_{i}}^{T}\exp (-E/k_{B}T^{\prime
})^{\prime }dT^{\prime }\right) .
\end{equation}

The thermoluminescence intensity $I$ as a function of temperature is given
as: 
\begin{equation}
I=-c\left( \frac{dn}{dt}\right) =cn_{0}s\left[ \exp (-E/k_{B}T)\right] \left[
\exp \left( -\frac{s}{r}\int_{T_{i}}^{T}\exp (-E/k_{B}T^{\prime })^{\prime
}dT^{\prime }\right) \right] .
\end{equation}

\section{Tsallis Thermostatistics}

Boltzmann-Gibbs statistics is used to study the systems having the following
conditions:\bigskip

(i) \ \ the spatial range of the microscopic interactions are short-ranged,

(ii) \ the time range of the microscopic memory is short-ranged,

(iii) the system evolves in a Euclidean-like space-time.\bigskip

These kinds of systems are said to be extensive. If a system does not obey
these restrictions, Boltzmann-Gibbs statistics seems to be inappropriate and
a non-extensive formalism must be used. Tsallis thermostatistics (TT) is one
of these formalisms.

TT has been applied to some concepts of thermostatistics [1] and also
achieved in solving some physical systems, where BG statistics is known to
fail: stellar polytropes [2], Levy-like anomalous diffusions [3-7],
two-dimensional Euler turbulence [8], solar neutrino problem [9], velocity
distributions of galaxy clusters [10], fully developed turbulence [11-14],
electron-positron and other high-energy collisions [15-19], anomalous
diffusion of Hydra viridissima [20] and nematic liquid crystals [21].\bigskip

Since the paper by Tsallis [22], there has been a growing tendency to
nonextensive statistical formalism. It has been shown that this formalism is
useful, because it provides a suitable theoretical tool to explain some of
the experimental situations, where standard thermostatistics seems to fail,
due to the presence of long-range interactions, or long-range memory
effects, or multi-fractal space-time constraints.

TT has been applied to various concepts of thermostatistics and achieved in
solving some physical systems, where Boltzmann-Gibbs statistics is known to
fail. Recently, Oliveira et al. [23] have pursued the idea, based on novel
experimental and theoretical results, that manganites are
magneticallynon-extensive objects. This property appears in systems where
long-range interactions and/or fractality exist, and such features have been
invoked in recent models of manganites, as well as in the interpretation of
experimental results. Therefore we could expect that this kind of materials
could be examined within a non-extensive thermostatistics, one of which is
called Tsallis thermostatistics (TT) and as you can see in our paper, TT has
been applied to various concepts of thermostatistics and achieved in solving
some physical systems, where Boltzmann-Gibbs statistics is known to fail. On
the other hand, we are still working on application of TT on
thermoluminescence process, so another studies on this subject are
forthcoming, experimentally and theoretically.

TT considers three possible choices for the form of a nonextensive
expectation value. These choices have been studied in [24] and applied to
two systems; the classical harmonic oscillator and the quantum harmonic
oscillator. In that study, Tsallis et al. studied three different
alternatives for the internal energy constraint. The first choice is the
conventional one,

\begin{equation}
\sum_{i=1}^{W}p_{i}\varepsilon _{i}=U_{q}^{(1)}.
\end{equation}
The second choice is given by

\begin{equation}
\sum_{i=1}^{W}p_{i}^{q}\varepsilon _{i}=U_{q}^{(2)}
\end{equation}
and regarded as the canonical one. Both of these choices have been applied
to many different systems in the last years [25]. However both of them have
undesirable difficulties. The third choice for the internal energy
constraint is

\begin{equation}
\frac{\sum_{i=1}^{W}p_{i}^{q}\varepsilon _{i}}{\sum_{i=1}^{W}p_{i}^{q}}%
=U_{q}{}^{(3)}.
\end{equation}
This choice is commonly considered to study physical systems because it is
the most appropriate one, and is denoted as the Tsallis-Mendes-Plastino
(TMP) choice. $q$ index is called the entropic index and comes from the
entropy definition,

\begin{equation}
S_{q}=k\frac{1-\sum_{i=1}^{W}p_{i}^{q}}{q-1}
\end{equation}

where $k$ is a constant, $\sum_{i}p_{i}=1$ is the probability of the system
in the $i$ microstate, $W$ is the total number of configurations. In the
limit $q\rightarrow 1$, the entropy reduces to the well-known
Boltzmann-Gibbs (Shannon) entropy.

\qquad The optimization of $S_{q}$ leads to

\begin{equation}
p_{i}^{(3)}=\frac{\left[ 1-(1-q)\beta (\varepsilon
_{i}-U_{q}^{(3)})/\sum_{j=1}^{W}(p_{j}^{(3)})^{q}\right] ^{\frac{1}{1-q}}}{%
Z_{q}^{(3)}}
\end{equation}

with

\begin{equation}
Z_{q}^{(3)}=\sum_{i=1}^{W}\left[ 1-(1-q)\beta (\varepsilon
_{i}-U_{q}^{(3)})/\sum_{j=1}^{W}(p_{j}^{(3)})^{q}\right] ^{\frac{1}{1-q}}.
\end{equation}
This equation is an implicit one for the probabilities $p_{i}$. Therefore
the $normalized$ $q-expectation$ $value$ of an observable is defined as

\begin{equation}
A_{q}=\frac{\sum_{i=1}^{W}p_{i}^{q}A_{i}}{\sum_{i=1}^{W}p_{i}^{q}}%
=\left\langle A_{i}\right\rangle _{q}
\end{equation}

where $A$ denotes any observable quantity which commutes with the
Hamiltonian. This expectation value recovers the conventional expectation
one when $q=1$. As mentioned above, Eq.(11) is an implicit one and in order
to solve this equation, Tsallis et al. suggest two different approaches; ''$%
iterative$ $procedure$'' and ''$\beta \rightarrow \beta ^{^{\prime }}$''
transformation.

\bigskip

The luminescence intensity of a pre-irradiated phosphor is recorded as a
function of the rising temperature. The resulting curve between the
luminescence intensity and temperature, known as the glow curve, first
increases and then, after reaching a maximum, decreases. If n represents the
ocuupied trap density at time t, thenthe following equation gives the rate
of thermal release of electron at temperature T within Tsallis
thermostatistics:

\begin{equation}
\frac{dn}{dt}=-ns(1-(1-q)E/kT)^{(\frac{1}{1-q})}.
\end{equation}

It is important to note that the rate is expressed as a power law rather
than exponential one. If the phosphor is heated at a constant rate of
increase of temperature, i.e. $dT/dt=cons\tan t=r,$ the above equation can
be rewritten as 
\begin{equation}
\frac{dn}{n}=-\frac{s}{r}(1-(1-q)E/kT)^{(\frac{1}{1-q})}.
\end{equation}

We assume that \ initially at the start of heating, $T=T_{i}$, $n=n_{0}$.
Then the integration of Eq.(13) results in 
\begin{equation}
n=n_{0}\left( 1+(1-q)\left( \frac{\left( -\frac{s}{r}%
\int_{T_{i}}^{T}(1-(1-q)E/kT^{\prime })^{(\frac{1}{1-q})}dT^{\prime }\right)
^{\frac{1}{1-q}}-1}{1-q}\right) \right) ^{\frac{1}{1-q}}.
\end{equation}

The thermoluminescence intensity $I$ as a function of temperature is given
as: 
\begin{eqnarray}
I &=&-c\left( \frac{dn}{dt}\right) =cn_{0}s\left[ (1-(1-q)E/kT)^{(\frac{1}{%
1-q})}\right]  \nonumber \\
&&\left( 1+(1-q)\left( \frac{\left( -\frac{s}{r}%
\int_{T_{i}}^{T}(1-(1-q)E/kT^{\prime })^{(\frac{1}{1-q})}dT^{\prime }\right)
^{\frac{1}{1-q}}-1}{1-q}\right) \right) ^{\frac{1}{1-q}}.
\end{eqnarray}

Figure shows the computed glow curves corresponding to Eq.(15). The
parameters in the Eq.(15) are: $q=5\,K.s^{-1}$, electron trap depth $%
E=0.8\,eV$ and $s=5\times 10^{11}s^{-1}$. In the figure, it is shown the
effect of the nonextensivity. In other words, if the rate of thermal release
of electrons at temperature T is expressed as a power law relation, then the
glow intensity vs temperature behaves as shown in the figure. An interesting
point is that in all the curves corresponding to different values of $q$
entropic index, the intensity approximately has the same value at $T\simeq
390K$.

\section{Results and Discussion}

Presumably there is some normalization at 390 K in order to compare with $q=1
$ case. The peak sequence moves smoothly with $q$ value, this is in much the
same way as caused by changing thermoluminescence kinetics if one varies the
relative number of trapping and recombination sites and includes secondary,
or back reactions. Whilst the data presented here are of a preliminary
nature, it is sufficient to give encouragement for further developments.

\newpage

\section*{\bf REFERENCES}

1. S. Abe, Y. Okamoto, Nonextensive statistical mechanics and its
applications, Series Lectures Notes in Physics, Berlin: Springer-Verlag,
2001.

2. A.R. Plastino, A. Plastino, Phys. Lett. A 174 (1993) 384.

3. P.A. Alemany, D.H. Zanette, Phys. Rev. E 49 (1994) R956.

4. C. Tsallis, S.V.F. Levy, A.M.C. Souza, R. Maynard, Phys. Rev. Lett. 75
(1995) 3589.

5. D.H. Zanette, P.A. Alemany, Phys. Rev. Lett. 75 (1995) 366.

6. M.O. Careres, C.E. Budde, Phys. Rev. Lett. 77 (1996) 2589.

7. D.H. Zanette, P.A. Alemany, Phys. Rev. Lett. 77 (1996) 2590.

8. B.M. Boghosian, Phys. Rev. E 53 (1996) 4754.

9. G. Kaniadakis, A. Lavagno, P. Quarati, Phys. Lett. B 369 (1996) 308.

10. A. Lavagno, G. Kaniadakis, M.R. Monteiro, P. Quarati, C. Tsallis,
Astrophys. Lett. Commun. 35 (1998) 449.

11. T. Arimitsu, N. Arimitsu, Phys. Rev. E 61 (2000) 3237.

12. T. Arimitsu, N. Arimitsu, J. Phys. A 33 (2000) L235.

13. C. Beck, Physic A 277 (2000) 115.

14. C. Beck, G.S. Lewis, H.L. Swinney, Phys. Rev. E 63 (2001) 035303.

15. D.B. Walton, J. Rafelski, Phys. Rev. Lett. 84 (2000) 31.

16. G. Wilk, Z. Wlodarcsyk, Nucl. Phys. B (Proc. Suppl.) 75A (1999) 191.

17. G. Wilk, Z. Wlodarcsyk, Phys. Rev. Lett. 84 (2000) 2770.

18. M.L.D. Ion, D.B. Ion, Phys. Lett. B 482 (2000) 57.

19. I. Bediaga, E.M.F. Curado, J. Miranda, Physica A 286 (2000) 156.

20. C. Beck, Physica A 286 (2000) 164.

21. A. Upadhyaya, J.-P. Rieu, J.A. Glazier, Y. Swada, Physica A 293 (2001)
549.

22. C. Tsallis, J. Stat. Phys. 52 (1988) 479.

23 M.S. Reis, J.P. Ar\'{a}{u}jo, V.S. Amaral, E.K. Lenzi and I.S. Oliveira,
Phys. Rev. B 66 (2002) 134417.

24. C. Tsallis, R.S. Mendes, A.R. Plastino, Physica A 261 (1998) 534.

25. see http://tsallis.cat.cbpf.br for updated bibliography.

\section*{\bf FIGURE CAPTIONS}

Figure. The computed glow curves corresponding to Eq.(15) for various values
of $q$.

\noindent

\end{document}